\documentclass[12pt] {article}
\usepackage{latexsym}{\rm }
\usepackage{graphics}
\usepackage{graphpap}
\newtheorem{lemma}{Lemma}
\newtheorem{theorem}{Theorem}
\newcommand{\beq}{\begin{equation}}
\newcommand{\feq}[1]{\label{#1} \end{equation}}
\newcommand{\beqr}{\begin{eqnarray}}
\newcommand{\feqr}{\end{eqnarray}}
\def\non{\nonumber}

\newcommand{\rf}[1]{(\ref{#1})}

\def\np#1#2#3{Nucl. Phys. {\bf{B#1}} (#2) #3}
\def\cm#1#2#3{Comm. Math. Phys. {\bf{#1}} (#2) #3}

\def\plb#1#2#3{Phys. Lett. {\bf{B#1}} (#2) #3}
\def\mpl#1#2#3{Mod. Phys. Lett. {\bf{A#1}} (#2) #3}

\begin{document}

\begin{center}


{\Large \bf Boundary States, Extended Symmetry Algebra and Module Structure for certain Rational Torus Models}\\ [4mm]

\large{Ioannis Smyrnakis} \footnote [2] 
{Email: smyrnaki@tem.uoc.gr}\\ [5mm]

{\small University of Crete, \\
Department of Applied Mathematics, \\
L. Knosou-Ambelokipi, 71409 Iraklio Crete,\\ 
Greece}
\vspace{5mm}

\end{center}

\begin{abstract}

The massless bosonic field compactified on the circle of rational $R^2$ is reexamined in the presense of boundaries.  
A particular class 
of models corresponding to $R^2=\frac{1}{2k}$ is distinguished by demanding the existence of a consistent set of 
Newmann boundary states. The boundary states are 
constructed explicitly for these models and the fusion rules are derived from them.  These are the ones prescribed by 
the Verlinde formula from the S-matrix of the theory.  In addition, the extended symmetry algebra of these theories is 
constructed which is responsible for the rationality of these theories.  Finally, the chiral space of these models is 
shown to split into a direct sum of irreducible modules of the extended symmetry algebra.   

\end{abstract}
\section{Introduction}
	The massless bosonic field compactified on the circle has been particularly useful for studying the moduli 
space of c=1 theories \cite{ginsparg} \cite{verlinde}.  
In this case, there are two continuous families 
of theories corresponding to the torus and the $Z_2$ orbifold models. Theories in each family are connected by marginal 
operator deformations.  In both cases there is a duality that identifies the model with radious R with the 
 model with radious $1/2R$.  Also the theory corresponding to $R=\sqrt{2}$ for the torus models is identified with 
the $R=1/\sqrt{2}$ theory for the orbifold models.  For the self-dual radious $R=1/\sqrt{2}$ 
the theory possesses an extended $SU(2)\times SU(2)$ symmetry.  By dividing this symmetry by the three special 
discrete subgroups of $SU(2)$, the tetrahedral, octahedral and icosahedral, \cite{ginsparg}
 constructed three more 
theories that are not connected to the others by marginal operator deformations.  This list of $c=1$ theories has 
been shown to be complete in the case the partition function is a linear combination of toroidal partition 
functions by \cite{verlinde}.
  However the partition functions of these theories contain an infinite 
sum of products of holomorphic times antiholomorphic Virasoro characters, as predicted by a theorem of \cite{cardy}.
  In the particular case $R^2$ is rational, 
these products group into a finite sum of holomorphic times antiholomorphic blocks.
  Furthermore this is the complete list of c=1 theories possesing this property as 
proven in \cite{kiritsis}.  
\par Next we consider conformal field theory on surfaces with boundaries \cite{alvarez}.
On the finite cylinder it is possible to construct a partition function in two ways.  Either through a 
closed string propagating between two boundary states or through an open string satisfying corresponding boundary 
conditions propagating around a loop.  Compatibility between these two points of view gives two conditions on 
the boundary states.  One is the Ishibashi condition \cite{ishibashi}
arising from the restriction of the conformal field theory to the upper half plane (necessary in the open string picture) 
and the other is the Cardy condition \cite{cardy1} \cite{petkova} 
which arises from the representational compatibility of the two constructions.  
Cardy in particular showed that given the Verlinde formula 
and a complete set of Ishibashi states it is possible to construct a consistent set of boundary states for diagonal 
theories.  In the particular case the representation space of the theory splits into a direct sum of irreducible 
representations of an 
extended $\hat{su}(2)$ chiral algebra the Ishibashi states have been constructed in \cite{ishibashi}.  
The case of the $Z_2$ orbifold model at radious 1 has been considered in \cite{affleck}.
In this case boundary states have been derived that do not correspond to bulk operators.   

\par In this work we study the rational torus models from the boundary CFT point of view.  We distinguish a 
particular class of models ($R=\frac{1}{\sqrt{2k}}$) which posesses a consistent set of Newmann boundary states (or 
dually Dirichlet boundary states) in the sense of \cite{cardy1}.
These boundary states have been constructed explicitly.  Next the extended symmetry algebra of these theories has 
been written down.  Finally the infinite direct sum of Fock modules, that constitutes the chiral representation space 
of this particular set of torus models, is shown to decompose into a finite sum of irreducible extended algebra modules.


\section{Torus Models Revisited}
\par These models correspond to a free massless real bosonic field compactified on a circle of radious R.  The 
action in this case is 
\beqr
S=\frac{1}{2\pi }\int \partial X \bar{\partial } X
\label{action}
\feqr
The partition function for these models turns out to be 
\beqr
Z(\beta )=\sum_{m,n} \frac{q^{\frac{1}{2}(\frac{m}{2R}+nR)^2}}{\eta (q)}
\frac{\bar{q}^{\frac{1}{2}(\frac{m}{2R}-nR)^2}}{\eta (\bar{q})}
\label{part}
\feqr
Here $q=e^{2\pi i\tau }=e^{-2\pi \beta }$, $\beta=-i\tau $, and $\eta(q)$ is the Jacobi eta function.  
The representation space for these models can be read of from the partition function to be 
\beqr
H=\bigoplus_{n,m\epsilon Z}F_{\alpha_{m,n}}\bigotimes F_{\bar{\alpha}_{m,n}}
\label{hilb}
\feqr
where the possible charges are $\alpha_{m,n}=\frac{m}{2R}-nR$ and $\bar{\alpha}_{m,n}=\frac{m}{2R}-nR$. 
\par In the particular case $R^2=\frac{p}{p'}$ the partition function becomes a finite sum of holomorphic times 
antiholomorphic parts:
\beqr
Z(q)=\sum_{r=0}^{2p-1}\sum_{s=0}^{2p'-1}f_{r,s}(q)f_{r,-s}(\bar{q})
\label{part1}
\feqr
where 
\beqr
f_{r,s}(q)=\sum_{n\epsilon Z} \frac{q^{2pp'(n+\frac{r}{4p}+\frac{s}{2p'})^2}}{\eta(q)}
\label{char}
\feqr
\par Now consider the free field more closely. We can view our field as describing 
an open string propagating along a strip periodic in the horizontal direction but with boundary conditions along the 
end point lines of the string.  What we really have in this way is an open string propagating along the sides of a 
cylinder, while it obeys boundary conditions on the end point circles.  The length of the circumference will be 
taken to be $\beta$, while the height of the cylinder will be 1/2.  This one loop open string diagram is equivalent to 
a closed string propagating along the cylinder from one boundary to the other.  The boundaries of the closed string 
are to be interpreted as states.  Following the closed string picture we will take time to be the vertical direction.  
The free field admits now the following expansion into oscillator modes:
\beqr
\phi(\sigma ,t)=\hat{x}+\frac{2\pi}{\beta}R\hat{w}\sigma +\frac{\pi}{\beta }\hat{p}t+\frac{1}{2}i\sum_{n\ne 0}
\left( \frac{a_n}{n}e^{-\frac{2\pi i}{\beta }n(t+\sigma )}+\frac{\tilde{a}_n}{n}e^{-\frac{2\pi i}{\beta }n(t-\sigma )}
\right) 
\label{expn}
\feqr
This splits into left and right modes according to 
\beqr
\phi(\sigma ,t)=\phi_L(x^+)+\phi_R(x^-)
\label{split}
\feqr
where $x^+=t+\sigma $ and $x^-=t-\sigma $.  Here 
\beqr
\phi_L(x^+)=\frac{\hat{x}}{2}+\frac{\pi}{\beta}a_0x^+ +\frac{1}{2}i\sum_{n\ne 0}\frac{a_n}{n}e^{-\frac{2\pi i}{\beta }nx^+} \non \\
\phi_R(x^-)=\frac{\hat{x}}{2}+\frac{\pi}{\beta}\tilde{a}_0x^- +\frac{1}{2}i\sum_{n\ne 0}
\frac{\tilde{a}_n}{n}e^{-\frac{2\pi i}{\beta }nx^-} 
\label{split1}
\feqr
where $a_0=\hat{p}/2+R\hat{w} $ and $\tilde{a}_0=\hat{p}/2-R\hat{w}$.   
The Hamiltonian in the closed string picture turns out to be 
\beqr
H_{\beta }=\frac{2\pi }{\beta}\left[(R\hat{w})^2+(\hat{p}/2)^2+\sum_{n=1}^{\infty }a_{-n}a_n+
\sum_{n=1}^{\infty }\tilde{a}_{-n}\tilde{a}_n-1/12\right]
\label{hamilt}
\feqr
Using the oscillator mode representation of the Virasoro generators 
\beqr
L_m=\frac{1}{2}\sum_{-\infty }^{\infty }:a_{m-n}a_n: \non \\
\bar{L}_m=\frac{1}{2}\sum_{-\infty }^{\infty }:\tilde{a}_{m-n}\tilde{a}_n: 
\label{sugawara}
\feqr
we have that 
\beqr
H_{\beta }=\frac{2\pi }{\beta }(L_0+\bar{L}_0-1/12)
\label{hamilt1}
\feqr
\par The boundary states of the closed string have to satisfy the Ishibashi condition \cite{ishibashi}:
\beqr
(L_n-\bar{L}_{-n})|B>=0
\label{ishibash}
\feqr
This is certainly satisfied if 
\beqr
(a_m\pm \tilde{a}_{-m})|B>=0
\label{ishibash1}
\feqr
The plus sign correspond to Newmann boundary conditions in the open string picture while the minus sign correspond 
to Dirichlet boundary conditions \cite{affleck}.  
Solving the conditions \rf{ishibash1} we get the following Ishibashi states:
\beqr
|\iota^N_{nR}>=e^{-\sum_{n=1}^{\infty}\frac{a_{-n}\tilde{a}_{-n}}{n}}|p=0,w=nR>\non \\
|\iota^D_{m/2R}>=e^{\sum_{n=1}^{\infty}\frac{a_{-n}\tilde{a}_{-n}}{n}}|\frac{p}{2}=\frac{m}{2R},w=0>
\label{ishibash2}
\feqr
Note that in the Newmann case the zero mode condition demands that there is no momentum while in the 
Dirichlet case it demands that there is no winding.  Now we have the following lemma:

\begin{lemma}
The Ishibashi states \rf{ishibash2} give rise to the following inner products:
\beqr
<\iota^N_{nR}|e^{-H_{\beta }/2}|\iota^N_{mR}>&=&\frac{\tilde{q}^{\frac{R^2n^2}{2}}}{\eta(\tilde{q})}\delta_{n,m} \non \\
<\iota^D_{n/2R}|e^{-H_{\beta }/2}|\iota^D_{m/2R}>&=&\frac{\tilde{q}^{\frac{n^2}{8R^2}}}{\eta(\tilde{q})}\delta_{n,m}\non \\
<\iota^N_{nR}|e^{-H_{\beta }/2}|\iota^D_{m/2R}>&=&\tilde{q}^{\frac{-1}{24}}\prod_{n=1}^{\infty }
\frac{1}{1+\tilde{q}^n} \delta_{n,0}  \delta_{m,0}=\frac{\sum_{n\epsilon Z}(-1)^n\tilde{q}^{n^2}}{\eta(\tilde{q})}
\delta_{n,0}  \delta_{m,0} 
\label{innprod}
\feqr
where $\tilde{q}=e^{-\frac{2\pi }{\beta }}$.  
\end{lemma}

The proof of this lemma is a simple free field calculation.  
Note that it is possible to interchange the Newmann and Dirichlet Ishibashi states by interchanging the directions 
of $t$ and $\sigma$.  
\par There is however another condition that must be satisfied by the boundary states \cite{cardy1}.  
This states that if $|X_A>$ are the boundary states, then 
\beqr
Z_{AB}(\beta )=<X_A|e^{-H_{\beta }/2}|X_B>=\sum_in_{AB}^i\chi_i(q)
\label{card}
\feqr
for some integers $n_{AB}^i$ where $q=e^{-2\pi i\beta}$. In particular there is a special vacuus state 
$|X_0>$ for which $n_{0j}^i=\delta_{i,j}$.  This condition arises from the possibility to see $Z_{AB}(\beta )$ 
as an one loop amplitude of an open string satisfying the boundary conditions A,B on the boundary circles.   
\par Now lets restrict ourselves to the Newmann sector.  Consider the case $R^2$ is rational, 
and lets demand that there is a Newmann vacuum boundary state 
\beqr
|X_0^N>=\sum_{n\epsilon Z}C_n|\iota^N_{nR}>
\label{newvac}
\feqr
Of course demanding a Dirichlet vacuum boundary state is completely equivalent because of the $t$, $\sigma $ 
interchange duality.  
Consider the partition function on the cylinder with two vacuum boundary states at the boundary circles:
\beqr
Z_{00}(\beta )=<X_0^N|e^{-H_{\beta }/2}|X_0^N>=\sum_{n\epsilon Z}|C_n|^2\frac{\tilde{q}^{\frac{R^2n^2}{2}}}{\eta(\tilde{q})}
\label{00part}
\feqr
According to the Cardy condition \cite{cardy1}, 
$Z_{00}(\beta )$ must be equal to the vacuum character.  
To change the variable from $\tilde{q}$ to q, we need to use the Poisson resummation formula, which in this 
case takes the form:
\beqr
\sum_{n\epsilon Z}\frac{\tilde{q}^{A(n+b)^2}}{\eta (\tilde{q})}=\frac{1}{\sqrt{2A}}\sum_{n\epsilon Z}
\frac{q^{\frac{n^2}{4A}}e^{2\pi inb}} {\eta(q)}
\label{poisson}
\feqr
In the case of the rational theories and in order to be able to get the identity character after use of formula \rf{poisson} 
 there must exist a minimum integer m such that the constants $C_n$ in the classes n=rmodm are equal.  This gives rise 
to the following partition function:
\beqr
Z_{00}(\beta )=\sum_{r=0}^{m-1}|C_r|^2\sum_{n\epsilon Z}\frac{\tilde{q}^{\frac{m^2R^2}{2}(n+\frac{r}{m})^2}}
{\eta(\tilde{q})}=\sum_{r=0}^{m-1}\frac{|C_r|^2}{mR}
\sum_{n\epsilon Z}\frac{q^{\frac{n^2}{2m^2R^2}}e^{2\pi in\frac{r}{m}}}
{\eta(q)}
\label{00part1}
\feqr
Now letting $n=mn'+s$ we get 
\beqr
Z_{00}(\beta )=\sum_{r,s=0}^{m-1}\frac{|C_r|^2}{mR}e^{2\pi irs/m}\sum_{n'\epsilon Z}
\frac{q^{\frac{(n'+s/m)^2}{2R^2}}}{\eta(q)}
\label{00part2}
\feqr
Since $s/m<1$, the power $q^{\frac{1}{2R^2}}$ appears only when $s=0$, and taking into account that in the vacuum 
character only integer powers of q appear and that $\sum_{r=0}^{m-1}\frac{|C_r|^2}{mR}\ne 0$ we have that 
\beqr 
R^2=\frac{1}{2k}.
\label{radious}
\feqr
These are the values of the radious that we are going to analyse further.  For these values the closed string 
partition function 
regroups to 
\beqr
Z(\beta )=\sum_{s=0}^{2k-1}\chi_s(q)\chi_{-s}(\bar{q})
\label{00part3}
\feqr
where 
\beqr
\chi_s(q)=\sum_{n\epsilon Z}\frac{q^{k(n+\frac{s}{2k})^2}}{\eta (q)}
\label{char1}
\feqr
The vacuum character must be a partial sum of $\chi_0(q)$.  Comparing this with $Z_{00}(\beta )$ we get that $s=0$, 
which in turn means that $\chi_0(q)$ itself is the vacuum character.  Furthermore it is necessary that all $|C_r|$ 
are equal, so that the value $s=0$ is the only one that survives the sum over r.  So, without loss of generality we 
can assume that m=1.  In this case we have 
\beqr
Z_{00}(\beta )=\frac{|C|^2}{R}\sum_{n\epsilon Z}\frac{q^{kn^2}}{\eta (q)}
\label{00part4}
\feqr
Equality with $\chi_0(q)$ demands that $|C|=\sqrt{R}=\frac{1}{\sqrt[4]{2k}}$.  So the corresponding vacuum state is 
\beqr
|X_0^N>=\frac{1}{\sqrt[4]{2k}}\sum_{n\epsilon Z} |\iota^N_{n/\sqrt{2k}}>=
\frac{1}{\sqrt[4]{2k}}\sum_{l=0}^{2k-1}\sum_{n\epsilon Z}|\iota^N_{\sqrt{2k}(n+l/2k)}>
\label{vac1}
\feqr
Now it is necessary (because of the Cardy condition) that 
\beqr
Z_{0m}(\beta )=\chi_m(q)=\frac{1}{\sqrt{2k}}\sum_{l=0}^{2k-1}e^{\frac{2\pi ilm}{2k}}\chi_l(\tilde{q})
\label{0mpart}
\feqr
This demands that the other boundary states are
\beqr
|X^N_m>=\frac{1}{\sqrt[4]{2k}}\sum_{l=0}^{2k-1}e^{\frac{2\pi ilm}{2k}}\sum_{n\epsilon Z}
|\iota^N_{\sqrt{2k}(n+l/2k)}>
\label{states}
\feqr
It is not too difficult to check that 
\beqr
Z_{m_1m_2}=\frac{1}{2k}\sum_{l,m=0}^{2k-1}e^{2\pi il\frac{(m_2-m_1+m)}{2k}}\chi_m(q)=\chi_{m_1-m_2}(q)
\label{mnpart}
\feqr
This is certainly an integer sum of characters so the Cardy condition is satisfied.  
It is worth mentioning that since the construction of the boundary states is based on the modular properties of the 
characters the Verlinde formula gives integer fusion rules as was necessary.  In particular we can read of from \rf{0mpart} 
that $S_l^j=\frac{1}{\sqrt{2k}}e^{\frac{2\pi ijl}{2k}}$ and from the Verlinde formula 
\beqr
\sum_{i=0}^{2k-1}S_i^jN^i_{k'l}=S_k'^jS_l^j/S_0^j
\label{verlinde}
\feqr
we can read the fusion rules  
\beqr
N^i_{k'l}=\delta_{(i,k'+l)mod2k}.
\label{fusion}
\feqr


\section{Extended Symmetry Algebra}
\par We will now try to find which extended symmetry gives rise to the characters $\chi_m(q) $ for each k.  
Consider first the identity character 
\beqr
\chi_0(q)=\frac{\sum_{n\epsilon Z}q^{kn^2}}{\eta(q)}=
q^{-1/24}\frac{1+2q^k+2q^{4k}+\cdots +2q^{n^2k}+\cdots}{(1-q)(1-q^2)\cdots (1-q^k)(1-q^{k+1})\cdots }
\label{idchare}
\feqr
Expanding in q we get
\beqr
&&\chi_0(q) = q^{-1/24}\Bigl( \left[1+P(1)q+P(2)q^2+\cdots +P(k-1)q^{k-1}\right]+ \non \\
&&+q^k\left[ [P(k)+2]+[P(k+1)+2P(1)]q+\cdots +[P(4k-1)+2P(3k-1)]q^{3k-1}\right]+ \cdots +\non \\
&&+q^{n^2k}\left[ [ P(n^2k)+\cdots +2]+
\cdots +[P((n+1)^2k-1)+\cdots +2P((2n+1)k-1)]q^{(2n+1)r-1}\right]+ \non \\
&&+\cdots \Bigr)
\label{idchare1}
\feqr
Here $P(n)$ is the number of partitions of n.  Now the Fock module built on the zero charge vacuum, $F_0$, has 
$P(n)$ linearly independent states at level n.  From the above expansion we see that at level k we have two extra 
linearly independent 
states that are orthogonal to 
$F_0$, and since they are the first such states they have to be killed by all $a_n$ for n positive.  So they 
can be taken to be highest weight Fock states of charge $\alpha_{nm}=\pm \sqrt{2k}$ because the $L_0$ eigenvalue $h_{nm}=k$. 
They are generated by the spin k currents
\beqr
J^{\pm}(z)=e^{\pm i\sqrt{2k}\phi(z)}
\label{currents}
\feqr
Here we have taken $\phi(z)=2\phi_L(x^+)$ and $z=e^{2\pi ix^+/\beta}$.  Clearly, $J^{\pm }$ adds charge $\pm \sqrt{2k}$ 
on the states it acts.  
There is of course another current, of spin 1, the $U(1)$ current which we take to be 
\beqr
J^0(z)=\frac{1}{\sqrt{2k}}i\partial\phi(z)
\label{0current}
\feqr

\par Now we have the following lemma:
\begin{lemma}
The modes of the above currents generate the space $\bigoplus_{n\epsilon Z}F_{n\sqrt{2k}}$.  
\end{lemma}
{\it Proof:}
\par To prove this consider the mode expansion of the $J^+(z)$ current acting on the zero charge vacuum:
\beqr
\sum_{n\epsilon Z}J^+_nz^{-n-k}|0>=e^{\sqrt{2k}\sum_{n>0}\frac{a_{-n}}{n}z^n}|\sqrt{2k}>
\label{cmodes}
\feqr
where by $|\sqrt{2k}>$ we denote the vacuum state of charge $\sqrt{2k}$. From this relation we see that 
\beqr
J^+_0|0>=0,\quad J^+_{-1}|0>=0,\quad \cdots \quad J^+_{-k+1}|0>=0\quad J^+_{-k}|0>=|\sqrt{2k}>.
\label{cmodes1}
\feqr
Similarly we get that $J_{-k}^-|0>=|-\sqrt{2k}>$.  These are the extra states at level k. Of course since the modes of 
$J^0(z)$ can act on those states all the Fock descendents of the states $|\pm \sqrt {2k}>$ are generated by the current 
modes.  If we move to level k+r then the number of them is $2P(r)$.  Together with the descendents of $|0>$ we have 
overall $P(k+r)+2P(r)$ states.  This takes account of the second bracket 
in \rf{idchare1}.  But at level 4k there are again two more states.  To take account of them consider the product
\beqr
J^+(z)J_{-k}^+|0>=J^+(z)|\sqrt{2k}>=z^{2k}e^{\sqrt{2k}\sum_{n>0}\frac{a_{-n}}{n}z^n}|2\sqrt{2k}> 
\label{cmodes2}
\feqr
This means that we have 
\beqr
J^+_0 J^+_{-k} |0>=0,\quad J^+_{-1}J^+_{-k}|0>=0,\quad \cdots \quad J^+_{-3k+1}J^+_{-k}|0>=0\quad 
J^+_{-3k}J^+_{-k}|0>=|2\sqrt{2k}>.
\label{cmodes 3}
\feqr
Considering also $J^-_{-3k}J^-_{-k}|0>$ we get the extra two states at level 4k.  
Again at level 4k+r we have all the Fock descendents of $|0>$, $J^\pm_{-k}|0>=|\pm \sqrt{2k}>$, and 
$J^\pm_{-3k}J^\pm_{-k}|0>=|\pm 2\sqrt{2k}>$.  The number of these states is $P(4k+r)+2P(3k+r)+2P(r)$.  
Continuing in this fashion we get two extra states appearing at every level of the form $n^2k$, as indicated by 
\rf{idchare}.  They are of the form 
\beqr
J^\pm_{-(2n-1)k}\cdots J^\pm_{-3k}J^\pm_{-k}|0>=|\pm n\sqrt{2k}> 
\label{stat1}
\feqr
Of course all the Fock descendents of these states are created by the $J^0$ modes so we have that the current 
modes generate completely the direct sum $\bigoplus_{n\epsilon Z}F_{n\sqrt{2k}}$.  This ends the proof of the lemma.  

\par The next question is what are the OPE's satisfied by these high spin currents.  This is not too difficult 
to obtain.  These are the following:
\beqr
J^0(z)J^{\pm }(w)&=&\pm \frac{1}{z-w}J^{\pm }(w)+reg \non \\
J^+(z)J^-(w)&=& \frac{1}{(z-w)^{2k}}expn_{O(2k-1)}e^{i\sqrt{2k}(\phi(z)-\phi(w))}+reg
\label{OPE}
\feqr
The second operator product have to be expressed in terms of the currents.  This is possible by expanding 
$\phi(z)$ near w:
\beqr
J^+(z)J^-(w) =\frac{1}{(z-w)^{2k}}expn_{O(2k-1)}e^{{2k}\left[(z-w)J^0(w)+\cdots +\frac{(z-w)^{2k-1}}{(2k-1)!}
\partial^{2k-2}J^0(w) \right]}+reg
\label{OPE1}
\feqr
Here by $expn_{O(2k-1)}$ we mean expansion up to and including terms of order $(z-w)^{2k-1}$.  
At this point it is worth mentioning that our product $J^+(z)J^-(w)$ is a bilocal field in the sense of \cite{todorov}.
There it was remarked that for k=1/2 (unaccepted value for us) this product is the generating function 
for the $W_{1+\infty}$ algebra.  Here however we will restrict our attention to integer values of k.
Now, the product \rf{OPE1} can be expanded in terms of Schur polynomials.  
In general the Schur polynomials are defined by 
\beqr
e^{\sum_{k=1}^{\infty}t_kz^k}=\sum_{N=0}^{\infty }z^NS_N(t_1,t_2,\cdots ,t_k,\cdots )
\label{schur}
\feqr
and they turn out to be 
\beqr
S_N(t_1,t_2,\cdots,t_k,\cdots)=\sum_{n_1,\cdots,n_k,\cdots}^{\sum_{k=1}^{\infty}kn_k=N}
\frac{t_1^{n_1}\cdots t_k^{n_k}\cdots}{n_1!\cdots n_k!\cdots }\label{schur1}
\feqr
  
Let us now define, following \cite{todorov}, 
the associated polynomials:
\beqr
f^{l}(\partial^iJ^0(z))=(l)!:S_{l}\left( J^0(z),\frac{\partial J^0(z)}{2!},\cdots,\frac{\partial^{l-1}J^0(z)}{l!}\right):
\label{biloc}
\feqr
They satisfy the following recurrence relation:
\beqr
f^{l+1}(\partial^iJ^0(z))=(J^0(z)+\partial)f^l(\partial^iJ^0(z))=\cdots=(J^0(z)+\partial )^lJ^0(z)
\label{recurr}
\feqr
\par Using now these new definitions we have 
\beqr
J^+(z)J^-(w) &=& \frac{1}{(z-w)^{2k}}\left(\sum_{N=0}^{2k-1}S_N\left(2kJ^0(w),\cdots,2k\partial^{2k-2}J^0(w)/(2k-1)!\right)
(z-w)^N \right)+reg= \non \\
&=& \sum_{N=0}^{2k-1}\frac{f^{N}(2k\partial^iJ^0(w))}{N!}(z-w)^{N-2k}+reg
\label{OPE2}
\feqr

\par The next question is what is the algebra satisfied by the modes of the currents $J^+(z),J^-(z)$.  This can be 
read of from the OPE's:
\beqr
[J^+_n,J^-_m]=\oint_0\frac{dw}{2\pi i}\oint_w \frac{dz}{2\pi i}z^{n+k-1}w^{m+k-1}J^+(z)J^-(w)
\label{comm}
\feqr
where the left integral is around the origin and the right is around w.  Expanding the product of currents as above 
we get 
\beqr
[J^+_n,J^-_m]=\sum_{N=0}^{2k-1}\frac{\Gamma (n+k)}{\Gamma(2k-N)\Gamma(N+n-k+1)}\oint_0\frac{dw}{2\pi i}w^{n+m+N-1}
\frac{f^N(2k\partial^iJ^0(w))}{N!}
\label{comm1}
\feqr
This suggests the following definition:
\beqr
V^k_{m+n,N}=\oint_0\frac{dw}{2\pi i}w^{n+m+N-1} f^N(2k\partial^iJ^0(w))=\oint_0\frac{dw}{2\pi i}w^{n+m+N-1}
(2kJ^0(w)+\partial)^{N-1}2kJ^0(w)
\label{nop}
\feqr
where we have made use of the recurrence relation \rf{recurr}. 
In this notation \rf{comm1} becomes 
\beqr 
[J^+_n,J^-_m]=\sum_{N=0}^{2k-1}\frac{\Gamma (n+k)}{\Gamma(2k-N)\Gamma(N+n-k+1)\Gamma(N+1)}V^k_{n+m,N}
\label{comm2}
\feqr
Following the same procedure for the remaining commutators we get
\beqr
[J^0_n,J^0_m]=\frac{n}{2k}\delta_{n+m} \qquad [J^0_n,J^{\pm}_m]=\pm J^{\pm}_{m+n}
\label{comm3}
\feqr
while the other commutators are 0.

\par Note that in particular if $k=1$ then $V^1_{m+n,0}=\delta_{m+n}$ and $V^1_{m+n,1}=2J^0_{m+n}$, so in this case 
the commutator \rf{comm2} becomes
\beqr
[J^+_n,J^-_m]=n\delta_{m+n}+2J^0_{m+n}
\label{comm4}
\feqr
giving rise to the $\hat{su}(2)$ current algebra.  
Nevertheless it is only for $k=1$ that this algebra closes so nicely.  In general the operators $V^k_{m+n}$ belong 
to the universal enveloping algebra of the U(1) current.


\section{Primary Fields}
\par Now that we have the extended symmetry algebra, the next question is what are the primary fields with 
respect to this algebra.  To answer this we need to consider the characters $\chi_m(q)$.  Under 
$q\rightarrow e^{2\pi i}q$ we pick a phase $e^{2\pi i\left(\frac{m^2}{4k}-\frac{1}{24}\right)}$.  So the 
conformal dimension of the corresponding field is $h_m=\frac{m^2}{4k}$, $0\le m\le 2k-1$.  
There are two Virasoro primary fields with this conformal dimension, the fields 
\beqr
V_m^{\pm }(z)=e^{\pm i\frac{m}{\sqrt{2k}}\phi(z)}.
\label{primaries}
\feqr
Observe that for m=2k we get the two currents.  
\par These primary fields satisfy the following operator product expansions:
\beqr
&&J^0(z)V_m^{\pm}(w)=\pm \frac{m}{2k}\frac{V_m^{\pm}(w)}{z-w}+reg \non \\
&&J^+(z)V^-_m(w)=\frac{1}{(z-w)^m}:V^+_{2k-m}(w)e^{2k(z-w)J^0(w)+2k\frac{(z-w)^2}{2!}\partial J^0(w)+\cdots 
+2k\frac{(z-w)^{m-1}}{(m-1)!}\partial^{m-2}J^0(w)}:+reg \non \\
&&J^-(z)V^+_m(w)=\frac{1}{(z-w)^m}:V^-_{2k-m}(w)e^{-2k(z-w)J^0(w)-2k\frac{(z-w)^2}{2!}\partial J^0(w)-\cdots 
-2k\frac{(z-w)^{m-1}}{(m-1)!}\partial^{m-2}J^0(w)}:+reg \non \\
&&
\label{pope}
\feqr
where the remaining OPE's are trivial.  Note that we can take the fields $V_m^+(z)=e^{i\frac{m}{\sqrt{2k}}\phi(z)}$, 
$0\le m\le 2k-1$ as a complete set of primary fields, since they are related to the $V_m^-(z)$ through the last of the 
relations \rf{pope}.

\section{Module Structure}
\par The primary fields $V_m^+(z)$ add charge $m/\sqrt{2k}$ to the vacuum so we  have 
\beqr
V_m^+(0)|0>=|\frac{m}{\sqrt{2k}}>
\label{vac}
\feqr
Recall now that the chiral Fock space associated with the partition function was 
$H=\bigoplus_{n,m\epsilon Z}F_{\alpha_{nm}}$, 
where $\alpha_{nm}=n\sqrt{2k}+\frac{m}{\sqrt{2k}}$.  The currents now can only add charges that are multiples of 
$\sqrt{2k}$.  So it is natural to decompose the space H into the sum 
\beqr
H=\bigoplus_{m=0}^{2k-1}{\cal H}_m
\label{hilb}
\feqr
where 
\beqr
{\cal H}_m=\bigoplus_{n\epsilon Z}F_{\frac{m}{\sqrt{2k}}+n\sqrt{2k}}
\label{irmod}
\feqr
is the space that is generated by the currents from $V_m^+(0)|0>=|\frac{m}{\sqrt{2k}}>$.  

\par Lets examine now more closely what is the action of the currents on ${\cal H}_m$.  Again by expanding 
the character $\chi_m(q)$ as a power series in q we see that there is one extra state appearing at the levels 
$n^2k-nm+m^2/4k$ and $n^2k+nm+m^2/4k$ for all positive integers n.  Considering the product $J^+(z)|m/\sqrt{2k}>$ we get 
\beqr
\sum_{n\epsilon Z}J^+_{-n}z^{n-k}|m/\sqrt{2k}>=z^m e^{\sqrt{2k}\sum_{n>0}\frac{a_{-n}}{n}z^n}|\sqrt{2k}+m/\sqrt{2k}>
\label{modem}
\feqr
This implies that 
\beqr
J^+_{-(k+m)}|m/\sqrt{2k}>=|\sqrt{2k}+m/\sqrt{2k}>
\label{stat1}
\feqr
and $J^+_{-r}|m/\sqrt{2k}>=0$ for all $r<k+m$.  Applying $J^+(z)$ on this new state a number of times we get 
eventually that 
\beqr
J^+_{-\left( (2n-1)k+m\right)}\cdots J^+_{-(3k+m)}J^+_{-(k+m)}|m/\sqrt{2k}>=|n\sqrt{2k}+m/\sqrt{2k}>
\label{stat2}
\feqr
These states account for the extra states at the levels $n^2k+nm+m^2/4k$.  Applying $J^-(z)$ similarly a number of times 
on the state $m/\sqrt{2k}$ we get 
\beqr
J^-_{-\left( (2n-1)k-m\right)}\cdots J^-_{-(3k-m)}J^-_{-(k-m)}|m/\sqrt{2k}>=|-n\sqrt{2k}+m/\sqrt{2k}>
\label{stat3}
\feqr
and these states account for the extra states at levels $n^2k-nm+m^2/4k$.  

\par Suppose now that the space ${\cal H}_m$ is reducible, as a module of our current algebra.  Then there must be 
a state $|v_m>$ other than $|\frac{m}{\sqrt{2k}}>$ such that 
\beqr
J^{\pm}_n|v_m>=0 \qquad n>0
\label{cond11}
\feqr
 and 
\beqr
J^0_n|v_m>=0 \qquad n>0.
\label{cond12}
\feqr
\rf{cond12} forces us to consider highest weight Fock states.  Such states in the module ${\cal H}_m$ are 
the states $|m/\sqrt{2k}+n\sqrt{2k}>$ for n integer.  Suppose now that $n>0$.  Applying $J^-(z)$ we get 
\beqr
\sum_{l\epsilon Z}J^-_lz^{-l-k}|m/\sqrt{2k}+n\sqrt{2k}>=z^{-(m+2nk)}e^{-\sqrt{2k}
\sum_{l>0}\frac{a_{-l}}{l}z^l}|m/\sqrt{2k}+(n-1)\sqrt{2k}>
\label{stat4}
\feqr
and this in turn implies that 
\beqr
J^-_{m+(2n-1)k} |m/\sqrt{2k}+n\sqrt{2k}> \ne 0
\label{stat5}
\feqr
Since $m+(2n-1)k$ is positive for positive n, $J^-_r|m/\sqrt{2k}+n\sqrt{2k}>$ cannot be 0 for all positive r.  
If now $n=-n'<0$ then we need to apply $J^+(z)$.  In this case we get 
\beqr
J^+_{-m+(2n'-1)k}|m/\sqrt{2k}-n'\sqrt{2k}>=|m/\sqrt{2k}-(n'-1)\sqrt{2k}>\ne 0
\label{stat6}
\feqr
Since $-m+(2n'-1)k$ is positive for positive $n'$ (negative n), $J^+_r|m/\sqrt{2k}+n\sqrt{2k}>$ 
cannot be 0 for all positive r.   So ${\cal H}_m$ cannot be reducible.  Hence we have the following theorem:

\begin{theorem}
The space $H=\bigoplus_{n,m\epsilon Z}F_{\alpha_{nm}}$ where $\alpha_{nm}=n\sqrt{2k}+\frac{m}{\sqrt{2k}}$ admits 
the decomposition $H=\bigoplus_{m=0}^{2k-1}{\cal H}_m$ into irreducible modules of the algebra generated by 
the modes of the spin k currents  $J^{\pm }(z)$ and the spin 1 current $J^0(z)$.  In terms of Fock modules we have  
${\cal H}_m=\bigoplus_{n\epsilon Z}F_{\frac{m}{\sqrt{2k}}+n\sqrt{2k}}$.
\end{theorem}


\section{Conclusions}
\par What we have achieved in this work is to single out a family of rational torus models by demanding the 
existense of a consistent set of Newmann (dually Dirichlet) boundary states.  For this models we have written down 
the extended symmetry algebra which restricts the number of blocks to a finite number.  Furthermore it is shown that 
the space of these torus models, which is an infinite sum of Fock modules, splits into a direct sum of a finite 
number of irreducible extended algebra modules.  
\par The extended symmetry algebra that has appeared is a W type algebra since it contains high spin currents.  It 
should be thought of as a generalization of the SU(2) current algebra at level one, a theory which corresponds to $k=1$ 
in our list of theories.  It is of interest to identify the algebras corresponding to general level, getting in this 
way theories that may not posess free field representations.  Some work in this direction has been done in the context of 
W-algebras by \cite{flohr}.  A study of the representation theory of such algebras may 
give new examples of rational conformal field theories.   
 
\bibliographystyle{plain}

\end{document}